\documentclass[paper a4]{article}
\title{Noncommutative relativistic particle.}
\author{A.A. Deriglazov\footnote{alexei@fisica.ufjf.br ~ On leave of
absence from Dept. Math. Phys., Tomsk Polytechnical University,
Tomsk, Russia.}}
\date{Dept. de Matematica, ICE, Universidade Federal de Juiz de Fora,\\
MG, Brasil.}
\begin{document}
\maketitle
\large
\begin{abstract}
Noncommutative version of D-dimensional relativistic particle is proposed.
We consider the particle interacting with the configuration space
variable $\theta^{\mu\nu}(\tau)$ instead of the numerical matrix. The
corresponding Poincare invariant action has a local symmetry, which
allows one to impose the gauge $\theta^{0i}=0, ~ \theta^{ij}=const$. The
matrix $\theta^{ij}$ turns out to be the noncommutativity parameter of
the gauge fixed formulation. Poincare transformations of the gauge
fixed formulation are presented in the manifest form.
\end{abstract}


\noindent
Noncommutative geometry [1, 2] in some nonrelativistic models arises
[3-7] as the result of canonical quantization [8, 9] of
underlying dynamical
systems with second class constraints. Nontrivial bracket for
the configuration space variables appears in this case as the Dirac bracket,
after taking into account the constraints presented in the model. In
particular, the noncommutative nonrelativistic particle in $D=1+2$ can be
obtained in this way starting from the action [6, 7]
\begin{eqnarray}\label{1}
S=\int dt\left[\dot x^av_a-\frac{1}{2m}v^2+
\dot v_a\theta_{ab}v_b\right].
\end{eqnarray}
where $x_a(t), ~ v_a(t)$ are the configuration space variables and
$\theta_{ab}=\bar\theta\epsilon_{ab}$.
It can be considered as the action of ordinary particle (with position
$x_a$) written in the first order form, with the
``Chern-Simons term'' for $v$ added: $\dot v\theta v$.
The physical sector consist of $x_a$ and the conjugated
momentum $p_a$. The Dirac bracket for $x_a$ turns out to be nontrivial,
with the noncommutativity parameter being $\bar\theta$ [7].

In this note we present and discuss noncommutative version for D-dimensional
relativistic particle. To start with, note that the Chern-Simons term can
be added to the first order action of the relativistic particle as well.
It do not spoil the reparametrization invariance. As a consequence,
the model will contain the desired relativistic constraint $p^2-m^2=0$.
The problem is
that the numerical matrix $\theta^{\mu\nu}$ do not respect the Lorentz
invariance. To resolve the problem, we consider a particle interacting
with a new configuration-space variable
$\theta^{\mu\nu}(\tau)=-\theta^{\nu\mu}(\tau)$,
instead of the constant matrix. The action constructed is manifestly
Poincare invariant and has local symmetry related with the variable
$\theta$. The last one can be gauged out, an admissible gauge is
$\theta^{0i}=0, ~ \theta^{ij}=const$. The noncommutativity parameter of the
gauge fixed version is then $\theta^{ij}$. As it usually happens in a
theory with local symmetries [10], Poincare invariance of the gauge
fixed version is combination of the initial Poincare and local
transformations which preserve the gauge chosen. In the case under
consideration, the resulting transformations turn out to be linear and
involve the constant matrix $\theta^{ij}$ (see Eqs.(\ref{18}) below).

Let us present details. The configuration space variables of the model
are $x^\mu(\tau), ~ v^\mu(\tau), ~ e(\tau), ~ \theta^{\mu\nu}(\tau)$,
with the Lagrangian action being
\begin{eqnarray}\label{2}
S=\int dt\left[\dot x^\mu v_\mu-\frac{e}{2}(v^2-m^2)+
\frac{1}{\theta^2}\dot v_\mu\theta^{\mu\nu}v_\nu\right].
\end{eqnarray}
Here $\theta^2\equiv\theta^{\mu\nu}\theta_{\mu\nu}, ~ \eta^{\mu\nu}=
(+,-, \ldots ,-)$. Insertion of the term $\theta^2$
in the denominator has the same meaning as for the veilbein in
the action of massless particle: $L=\frac{1}{2e}\dot x^2$. Technically,
it rules out the degenerated gauge $e=0$. The action is manifestly
invariant under the Poincare transformations
\begin{eqnarray}\label{3}
x'^\mu=\Lambda^\mu{}_\nu x^\nu+a^\mu, \quad v'^\mu=\Lambda^\mu{}_\nu
v^\nu, \quad
\theta '^{\mu\nu}=\Lambda^\mu{}_\rho\Lambda^\nu{}_\sigma
\theta^{\rho\sigma}.
\end{eqnarray}
Local symmetries of the model are reparametrizations (with
$\theta^{\mu\nu}$ being the scalar variable), and the following
transformations with the parameter $\epsilon_{\mu\nu}(\tau)=-
\epsilon_{\nu\mu}(\tau)$
\begin{eqnarray}\label{4}
\delta x_\mu=v^\nu\epsilon_{\nu\mu}, \qquad
\delta\theta_{\mu\nu}=-\theta^2\epsilon_{\mu\nu}+
2\theta_{\mu\nu}(\theta\epsilon).
\end{eqnarray}
To analyse the physical sector of this constrained system, we rewrite
it in the Hamiltonian form.
Starting from the action (\ref{2}), one finds in the Hamiltonian
formalism the primary constraints
\begin{eqnarray}\label{5}
G^\mu\equiv p^\mu-v^\mu=0, \qquad
T^\mu\equiv\pi^\mu-\frac{1}{\theta^2}\theta^{\mu\nu}v_\nu, \cr
p_{\theta}^{\mu\nu}=0, \qquad \qquad p_e=0
\end{eqnarray}
and the Hamiltonian
\begin{eqnarray}\label{6}
H=\frac{e}{2}(v^2-m^2)+\lambda_{1\mu}G^\mu+\lambda_{2\mu}T^\mu+
\lambda_ep_e+\lambda_{\theta\mu\nu}p_{\theta}^{\mu\nu}.
\end{eqnarray}
Here $p, ~ \pi$ are conjugated momentum for $x, ~ v$ and $\lambda$ are the
Lagrangian multipliers for the constraints. On the next step there is
appear the secondary constraint
\begin{eqnarray}\label{7}
v^2-m^2=0,
\end{eqnarray}
as well as equations for determining the Lagrangian multipliers
\begin{eqnarray}\label{8}
\lambda_2^\mu=0, \qquad
\lambda_1^\mu=ev^\mu+\frac{2}{\theta^2}(\lambda_\theta v)^\mu-
\frac{4}{\theta^4}(\theta\lambda_{\theta})(\theta v)^\mu.
\end{eqnarray}
There is no of tertiary constraints in the problem.
Equations of motion follow from (\ref{6})-(\ref{8}), in particular,
for the variables $x, ~ p$ one has
\begin{eqnarray}\label{20}
\dot x^\mu=ep^\mu+\frac{2}{\theta^2}(\lambda_\theta v)^\mu-
\frac{4}{\theta^4}(\theta\lambda_{\theta})(\theta v)^\mu, \qquad
\qquad \dot p^\mu=0
\end{eqnarray}
Poisson brackets of the constraints are
\begin{eqnarray}\label{9}
\{G^\mu, G^\nu\}=0, \qquad \qquad \{T^\mu, T^\nu\}=
-\frac{2}{\theta^2}\theta^{\mu\nu}, \cr
\{G_\mu, T^\nu\}=-\delta_\mu^\nu, \qquad
\{T_\mu, p_\theta^{\rho\sigma}\}=-\frac{1}{\theta^2}
\delta_\mu^{[\rho}v^{\sigma ]}+
\frac{4}{\theta^4}(\theta v)_\mu\theta^{\rho\sigma}.
\end{eqnarray}
The constraints $G^\mu, ~ T^\mu$ form the second class subsystem and can
be taken into account by transition to the Dirac bracket. Then the
remaining constraints can be classified in accordance with their properties
relatively to the Dirac bracket. Consistency of the procedure is
guaranteed by the known theorems [9]. Introducing the Dirac bracket
\begin{eqnarray}\label{10}
\{A, B\}_D=\{A, B\}+2\{A, G_\mu\}\frac{1}{\theta^2}\theta^{\mu\nu}
\{G_\nu, B\}- \cr
\{A, G^\mu\}\{T_\mu, B\}-\{A, T_\mu\}\{G^\mu, B\},
\end{eqnarray}
one finds, in particular, the following brackets for the
fundamental variables (all the nonzero brackets are presented)
\begin{eqnarray}\label{11}
\{x^\mu, x^\nu\}=-\frac{2}{\theta^2}\theta^{\mu\nu}, \quad
\{x^\mu, p_\nu\}=\delta^\mu_\nu,
\quad \{p_\mu, p_\nu\}=0;
\end{eqnarray}
\begin{eqnarray}\label{12}
\{x^\mu, v_\nu\}=-\delta^\mu_\nu, \quad
\{x^\mu, \pi^\nu\}=-\frac{1}{\theta^2}\theta^{\mu\nu}, \quad
\{\theta_{\mu\nu}, p_\theta^{\rho\sigma}\}=-\delta_\mu^{[\rho}
\delta_\nu^{\sigma ]}, \cr
\{x^\mu, p_\theta^{\rho\sigma}\}=-\{\pi^\mu, p_\theta^{\rho\sigma}\}=
\frac{1}{\theta^2}\eta^{\mu [\rho}v^{\sigma ]}-
\frac{4}{\theta^4}(\theta v)^\mu\theta^{\rho\sigma}.
\end{eqnarray}
Let us choose $x^\mu, ~ p^\mu$ as the physical sector variables (one can
equivalently take $(x, v)$ or $(x, \pi)$, which leads to the same final
results, similarly to the nonrelativistic case [7]).
The variables $v, \pi$ can be omitted now from the consideration.

Up to now the procedure preserves the manifest Poincare invariance of
the model. Let us discuss the first class constraints
$p_\theta^{\rho\sigma}=0$. As the gauge fixing conditions one takes
\begin{eqnarray}\label{13}
\theta^{0i}=0, \qquad \qquad \theta^{ij}=const.
\end{eqnarray}
Then $\theta^{\mu\nu}\theta_{\mu\nu}=
-\theta_{ij}\theta_{ji}$, and the gauge is admissible if
$\theta_{ij}\theta_{ji}\ne 0$, see Eq.(\ref{11}), (\ref{20}). From the
equation of motion
$\dot\theta=\lambda_\theta$ one determines the remaining Lagrangian
multipliers: $\lambda_\theta=0$. Using this result in Eq.(\ref{20}),
the final form of the equations of motion is
\begin{eqnarray}\label{14}
\dot x^\mu=ep^\mu, \qquad \dot p^\mu=0.
\end{eqnarray}
They are supplemented by the
remaining first class constraints $p^2-m^2=0, ~ p_e=0$. Brackets for
the physical variables are given by the Eqs.(\ref{11}).

The initial Poincare transformations (\ref{3}) do not preserve the gauge
(\ref{13}) and must be accompanied by compensating local transformation,
with the parameter $\epsilon^{\mu\nu}$ chosen in appropriate way.
It gives the Poincare symmetry of the gauge fixed version. To find it,
one has the conditions ($\Lambda^\mu{}_\nu=\delta^\mu_\nu+
\omega^\mu{}_\nu$)
\begin{eqnarray}\label{15}
(\delta_\omega+\delta_\epsilon)\theta^{0i}=\omega^0{}_j\theta^{ji}+
\theta^2\epsilon^{0i}=0, \cr
(\delta_\omega+\delta_\epsilon)\theta^{ij}=\omega^{[i}{}_k\theta^{kj]}
\theta^2\epsilon^{ij}+2\theta^{ij}(\theta^{kp}\epsilon_{kp})=0
\end{eqnarray}
The solution is
\begin{eqnarray}\label{16}
\epsilon^{0i}(\omega)=\frac{1}{\theta^2}\omega^{0j}\theta^{ji}, \quad
\epsilon^{ij}(\omega)=\frac{1}{\theta^2}\omega^{[i}{}_k\theta_k{}^{j]},
\end{eqnarray}
or, equivalently
\begin{eqnarray}\label{17}
\epsilon^{\mu\nu}(\omega)=\frac{1}{\theta^2}
\omega^{[\mu}{}_\rho\theta^{\rho\nu]},
\end{eqnarray}
where Eq.(\ref{13}) is implied. Then the Poincare transformations of
the gauge fixed version are
\begin{eqnarray}\label{18}
\delta x^\mu=\omega^\mu{}_\nu x^\nu+\frac{1}{\theta^2}p_\nu
\omega^{[\nu}{}_\rho\theta^{\rho\mu]}, \quad
\delta p^\mu=\omega^\mu{}_\nu p^\nu.
\end{eqnarray}

\end{document}